\documentclass[doublecol,amsmath,amssymb,cite]{epl2}

\title{Couette Flow of Two-Dimensional Foams}
\shorttitle{} %Insert here a short version of the title if it exceeds 70 characters

\author{Gijs Katgert\inst{1} \and Brian P. Tighe\inst{2} \and Matthias E. M\"{o}bius\inst{1} \and Martin van Hecke\inst{1}}
\shortauthor{G. Katgert \etal}

\institute{ \inst{1}
  Kamerlingh Onnes Laboratorium, Universiteit Leiden,
   P.O. Box 9504, 2300 RA Leiden, The Netherlands. \\

\inst{2} Instituut Lorentz, Universiteit Leiden,
   P.O. Box 9506, 2300 RA Leiden, The Netherlands. \\}



 %rheo of


\abstract{We experimentally investigate flow of quasi
two-dimensional disordered foams in Couette geometries, both for
foams squeezed below a top plate and for freely floating foams
(bubble rafts). With the top plate, the flows are strongly
localized and rate dependent. For the bubble rafts the
flow profiles become essentially rate-independent, the local and
global rheology do not match, and in particular the foam flows in
regions where the stress is below the global yield stress. We
attribute this to nonlocal effects and show that the  ``fluidity''
model recently introduced by Goyon {\em et al.} ({\em Nature},
{\bf 454} (2008)) captures the essential features of flow both
with and without a top plate.}

\begin{document}

\maketitle
\section{Introduction}
Foams have recently attracted attention as model systems for
disordered, complex fluids \cite{cohen}. The elementary building
blocks, the bubbles, obey simple laws: when compressed, their
repulsion is harmonic \cite{lacasse, princen} and when sliding
past other bubbles or boundaries, they experience a velocity
dependent drag force \cite{bretherton, quere, denkov1, denkov2,
terriac, denkov3, drenckhan}.

Collectively, the conglomerate of bubbles that makes up a foam
exhibits all the hallmarks of complex fluids
--- foams exhibit shear banding, a yield stress  and
shear-thinning \cite{larson}. The latter are often modeled by a
Herschel-Bulkley constitutive equation where the stress $\tau$
takes the form $\tau = \tau_y+k \dot{\gamma}^{\beta}$, where
$\tau_y, k, \beta$ and $\dot{\gamma}$ denote the yield stress,
consistency, power law index and strain rate \cite{larson}.

The rheology of foams has mainly been investigated in three
dimensions \cite{cohen}. While the three-dimensional case is
perhaps more realistic, the opacity of foams inhibits connecting
the bulk behavior with the local behavior. Therefore, recently a
body of work has focused on the shear flow of two dimensional
foams. In this case the bulk response can easily be connected to
local quantities such as velocity profiles and bubble fluctuations
\cite{katgert, mobius, denninPRE06}.

The flow of two dimensional foams has  been studied extensively in
Couette geometries. For example, Dennin and co-workers have
sheared (freely floating) bubble rafts in a Couette geometry with
a fixed inner disk and a rotating outer cylinder
\cite{denninprl89, denninprl93}, while Debr\'{e}geas has confined
foam bubbles in a Hele-Shaw cell and rotated the inner disk,
keeping the outer cylinder fixed \cite{debregeasprl87}. In both
cases, localized flow profiles were found. There has
been no clear consensus, however, on the cause of flow
localization in these systems: while the (geometry induced) decay
of the stress away from the inner cylinder may cause
localization, a confining glass boundary, by introducing
additional drag forces on the foam, can have the same effect
\cite{katgert,denninpre73,krishan,clancy}. Thus the question of
what causes flow localization remains, and differing
opinions abound in the community \cite{cheddadi, debregeasjfm,
denninreview}.

In this Letter, we address this question by combining
measurements of the flow profiles of two-dimensional disordered
foams in Couette geometries with and without a top plate and for a
wide range of driving rates with independent rheological
measurements. When the top plate is present the flows are rate
dependent, while the flow profiles become essentially
rate-independent in absence of this plate --- in both cases, the
flow profiles are localized.

A recent model, developed by Janiaud et al.~and Katgert
et al.~successfully predicts velocity profiles in {\em linearly}
sheared monolayers \cite{katgert, janiaud} by  balancing a
Herschel-Bulkley expression for the stress with a drag force due
to the top plate. Here we show that this model
unexpectedly breaks down in Couette geometries \cite{clancy}.
This is seen most dramatically in the case without a top plate,
where, due to the presence of a yield stress, Herschel-Bulkley
rheology predicts a range of driving rates for which the flow
velocity vanishes at a point {\em within} the gap. In these cases,
we observe instead a velocity profile that decays to zero only at
the outer boundary. Thus, the material appears to flow below the
yield stress! Moreover, the Herschel-Bulkley constitutive relation
fails to capture the observed rate independence.

Thus, the Herschel-Bulkley model fails to capture the
flow, and we show that this breakdown reflects nonlocal flow
behavior. By applying the nonlocal model recently introduced by
Goyon {\em et al.} \cite{goyon,bocquet}, which contains a finite
cooperativity length $\xi$  that encodes the spatial extent of
plastic rearrangements, we find that we can describe all features
of the flow of foam in Couette geometry.

\begin{figure}[tb]
\centering
\includegraphics[width=0.9\columnwidth]{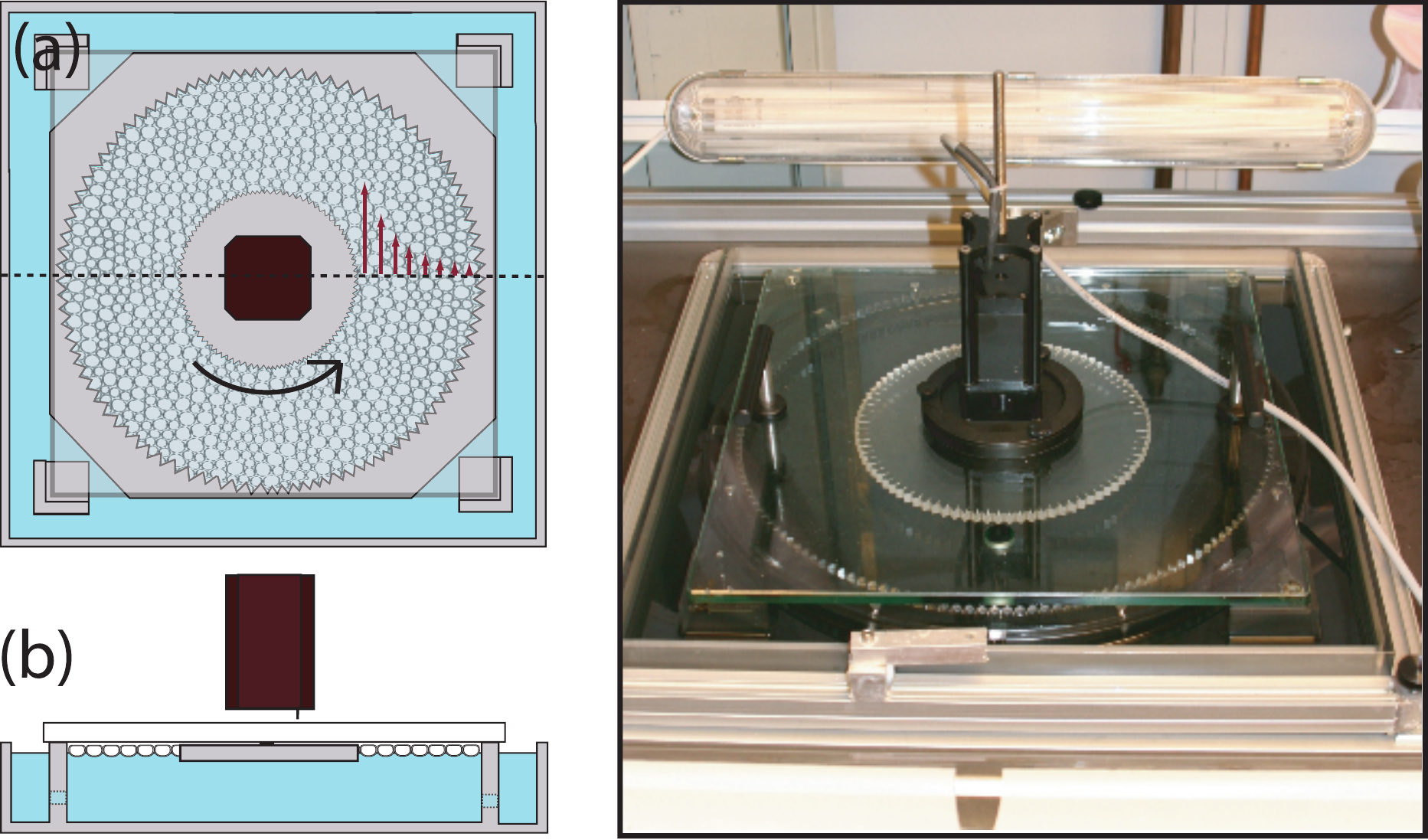}
\caption{(a) Schematic top view of one Couette cell used in this
experiment. The inner disk has radius $r_i = 105$ mm and the gap
has width $85$ mm. The outer cylinder, reservoirs and supports for
the glass plate have been milled into a PMMA block. (b) Side view:
the reservoirs and the bounded area are connected to keep the
region underneath the glass plate from draining. The motor is
connected to the inner cylinder through the glass plate. (c)
Photograph of the experimental setup.}   \label{couettesetup}
\end{figure}

\section{Experiment}
Our experimental setup consists of a $500 \times 500 \times 50$ mm
square PMMA block, into which the outer cylinder, a reservoir and
supports for a removable glass plate are milled, see
Fig.~\ref{couettesetup}. The boundary of the reservoir acts as the
outer cylinder (of radius $r_o =$ 190 mm) and is grooved with 6 mm
grooves. On the glass plate of $405 \times 405 \times 12$ mm, a
casing for a stepper motor is fixed by UV curing glue. The stepper
motor (L-5709 Lin engineering) is connected to an inner cylinder
of $r_i =$ 105 mm radius through a hole in the glass plate. The
inner cylinder is grooved like the outer cylinder.
\par
A bidisperse foam is produced by filling the reservoir with a surfactant
solution, the details of which can be found in \cite{katgert} and bubbling nitrogen through the fluid (viscosity
$\eta=1.8\pm0.1$ ${\rm mPa}\cdot{\rm s}$ and surface tension
$\sigma=28\pm 1$ mN/m) \cite{katgert}. After thorough mixing, we
obtain a bidisperse, disordered foam monolayer. The resulting
bubble sizes are  $d_1, d_2 = 1.8,~2.7$ mm. The glass plate, with
the inner driving wheel attached, is carefully placed on top of
the foam and subsequently, the foam is allowed to equilibrate for
a considerable time. Approximately 40 bubble layers span the
distance between the inner wheel and the outer cylinder.
To perform bubble raft experiments, we place
spacers between the supports and the glass plate, thus obtaining
a considerable gap between the liquid surface and the glass
plate.
\par
The foam is lit laterally by 4 fluorescent tubes and images are
recorded by a CCD camera (Foculus FO 432BW), equipped with a
Tamron 280-300 telezoom lens. The bottom of the reservoir is black
to enhance contrast. The frame rate is fixed  such that the
angular displacement of the inner cylinder is fixed at 1.12$\times
10^{-3}$ rad/frame. We record only during steady shear, ensuring
that the foam has been sheared considerably before starting image
acquisition.
\par
\begin{figure}[t]
\begin{center}
\includegraphics[width=0.9\linewidth]{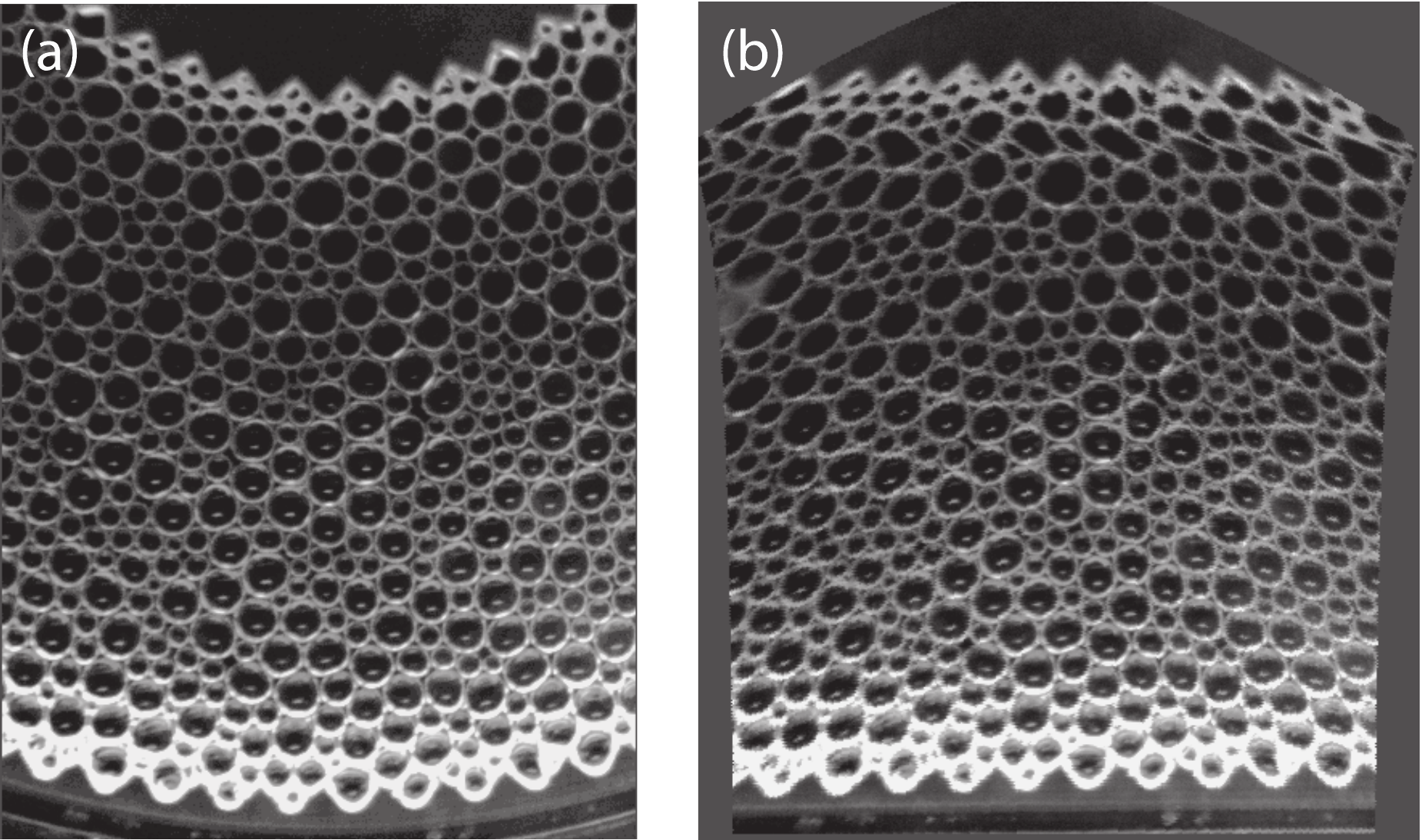}
\end{center}
\caption{(Color online) (a) Raw image as obtained by CCD camera.
The local curvature is extracted from the curvature at the inner
disk and the outer cylinder, and for every $r$ we define an arc
that we match to pixels in the image. If we plot these arcs as
straight lines we obtain: (b) the image with correction for
curvature.  We compute cross correlations between subsequent
frames on these straightened image lines. }   \label{curved}
\end{figure}
\begin{figure}[t]
\begin{center}
\includegraphics[width=0.9\linewidth]{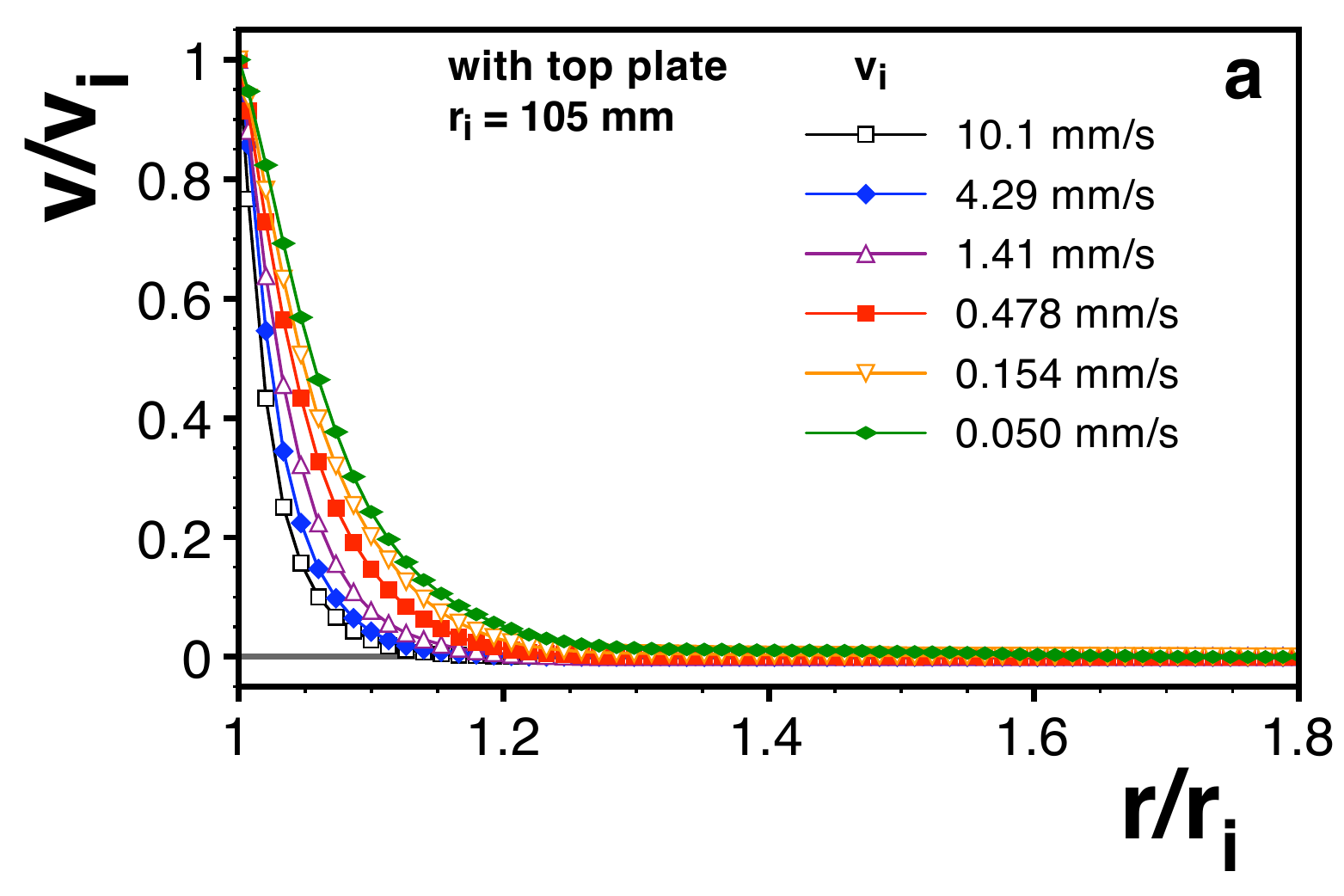}\\
\includegraphics[width=0.9\linewidth]{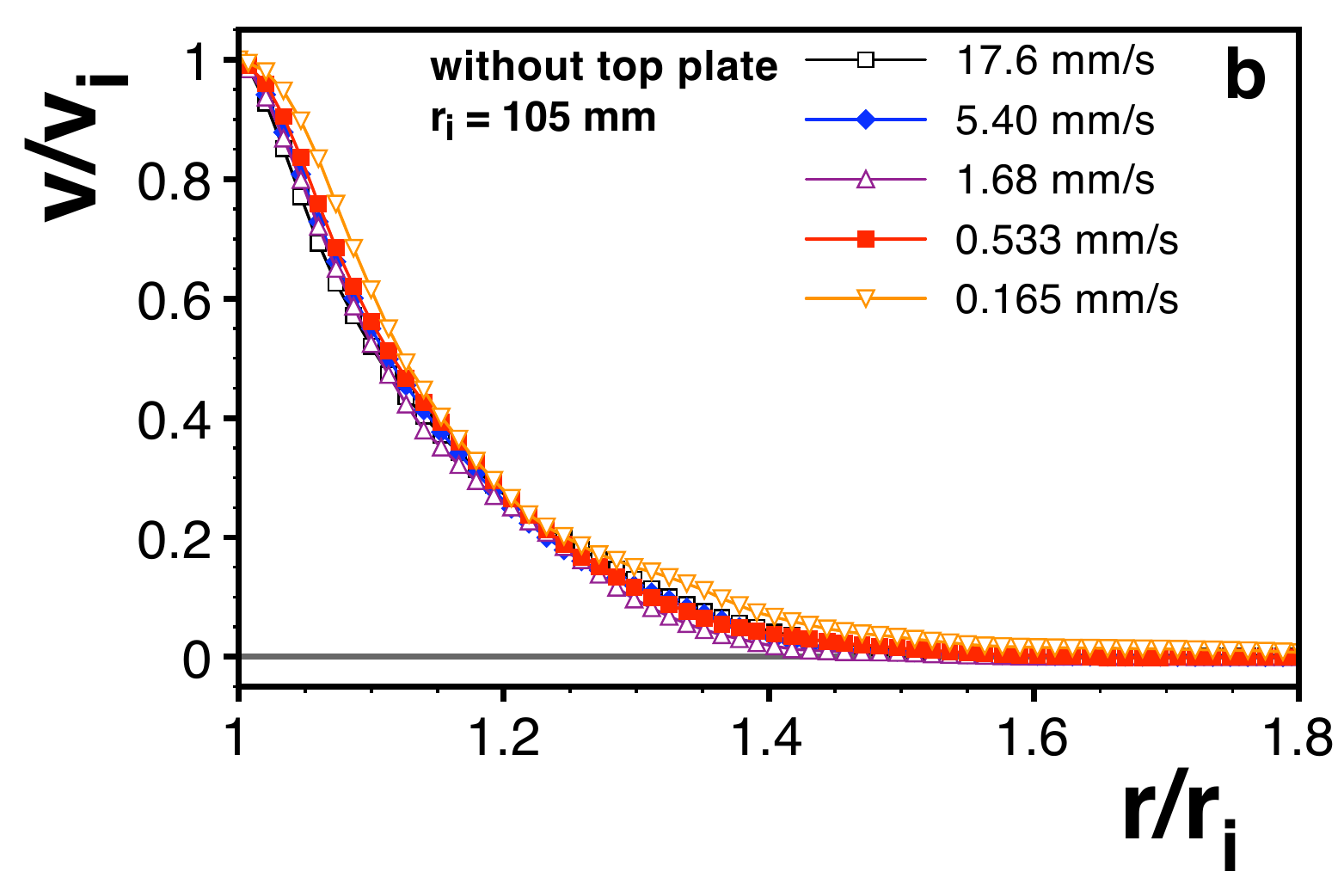}
\end{center}
\caption{(Color online) (a) Velocity profiles for two dimensional
Couette flow of foam with top plate. We see strongly
localized velocity profiles that furthermore exhibit
rate dependence: the faster the driving velocity $v_i$,
the more localized the profiles become. (b) Velocity
profiles for two dimensional Couette flow of foam without top
plate. We see approximately rate independent velocity profiles,
with localization that is solely due to the curved geometry.}
\label{profiles}
\end{figure}
We calculate velocity profiles across the gap between inner and
outer wheel by cross-correlating arcs of fixed radial distance in
subsequent frames over a large angular region. This approach
forces us to calculate velocity profiles on curved image lines.
However, by defining circular arcs and identifying these with the
appropriate pixels, this can easily be done, see
Fig.~\ref{curved}(a,b). We compute averaged velocities over
between 2,000 and 10,000 frames, depending on the experiments,  to
enhance statistics. We assume the accuracy of the profiles to be
at best 0.001 pixel/frame and cut off velocities that fall below
this threshold. We check that coarsening, segregation, coalescence
and rupturing are absent in the runs with a top plate, whereas we
do observe rupturing in the bubble raft experiment:
bubbles will pop after approximately $1 \frac{1}{2}$
hours. We merely content ourselves with the absence of holes in
our foam during the latter experiment, which
is achieved by loading the Couette cell with a surplus
of foam far away from the imaging region.

\section{Results}

In Fig.~\ref{profiles}(a) we present data for the shear
flow of a foam covered with a glass plate, at 6 different driving
velocities $v_i:=v(r_i)$ spanning 2.5 decades. We have rescaled
the velocity profiles with $v_i$ to highlight the qualitative
changes and we have rescaled the radial coordinate with the inner
radius $r_i$, which characterizes the curvature of the
experimental geometry and, in the case without a top plate, sets
the decay of the stress profile. We observe that the shape of the
velocity profiles depends on the exerted rate of strain; the runs
that were recorded at the highest driving velocity exhibit the
most localization.

The observed rate dependence is in accordance with our previous
results \cite{katgert}, obtained for the linear shear of
two-dimensional foams bounded by a top plate, which also displayed
rate dependent localization in the presence of a top
plate. From these results, we can infer that again the
balance between internal dissipation in the foam and the external
top plate drag force leads to increased localization at increasing
shear rates. Note that our results are in strong contrast with
the findings by Debr\'{e}geas et al. \cite{debregeasprl87}, where
rate independent profiles were found. As in \cite{debregeasprl87},
the decay of the velocity profile is approximately exponential
(semi-log plot not shown).

We now turn to Couette shear of a freely flowing foam
without drag from the top plate.  Despite the limited
stability of the bubble raft, we can shear the foam at the same
shear rates as in the experiments with a bounding glass plate,
except for the slowest run. Results are plotted in
Fig.~\ref{profiles}(b): within experimental uncertainty the
profiles exhibit rate independent velocity profiles. We observe
that the velocity profiles are still reasonably
localized.

Localization absent the top plate is not itself
surprising: Herschel-Bulkley  fluids are expected to display
localization in Couette flow when the shear stress,
which decays as $1/r^2$, falls below the yield stress. At this
point, the local strain rate is discontinuous. Here we do not
observe any such discontinuity in the local strain rate. This
should be contrasted to earlier claims of flow discontinuities by
Rodts et al.\cite{rodts}~and Dennin and co-workers
\cite{denninprl93, denninreview, denninpre74} experimentally and
Cheddadi et al.~theoretically \cite {cheddadi}. We note here that
the experimental proof of a flow discontinuity is highly non-trivial 
--- that is why we turn our attention to rheometry.
We have not seen packing density gradients associated with the flow
localization, and gradients are smooth enough (except very close to 
the wall), to exclude effects due to the discrete nature of the 
material \cite{isa}.

%\revision{ Localization} absent the top plate is not itself
%surprising: Herschel-Bulkley and power law fluids are expected to
%display \revision{localization} in Couette flow when the gap is
%sufficiently wide that the $1/r^2$ decay of the stress cannot be
%neglected. We observe, however, no \revision{discontinuity in the
%local strain rate, which would indicate} a flowing and a static
%region. This is expected for a Herschel-Bulkley fluid (see below),
%and was found by Rodts et al.\cite{rodts}~and Dennin and
%co-workers \cite{denninprl93, denninreview, denninpre74}
%experimentally and Cheddadi et al.~theoretically \cite {cheddadi}.
%\revision{Note that the experimental proof of a flow discontinuity
%in foams is controversial.}

\section{Rheometry}
\begin{figure}[t]
\begin{center}
\includegraphics[width=0.9\linewidth]{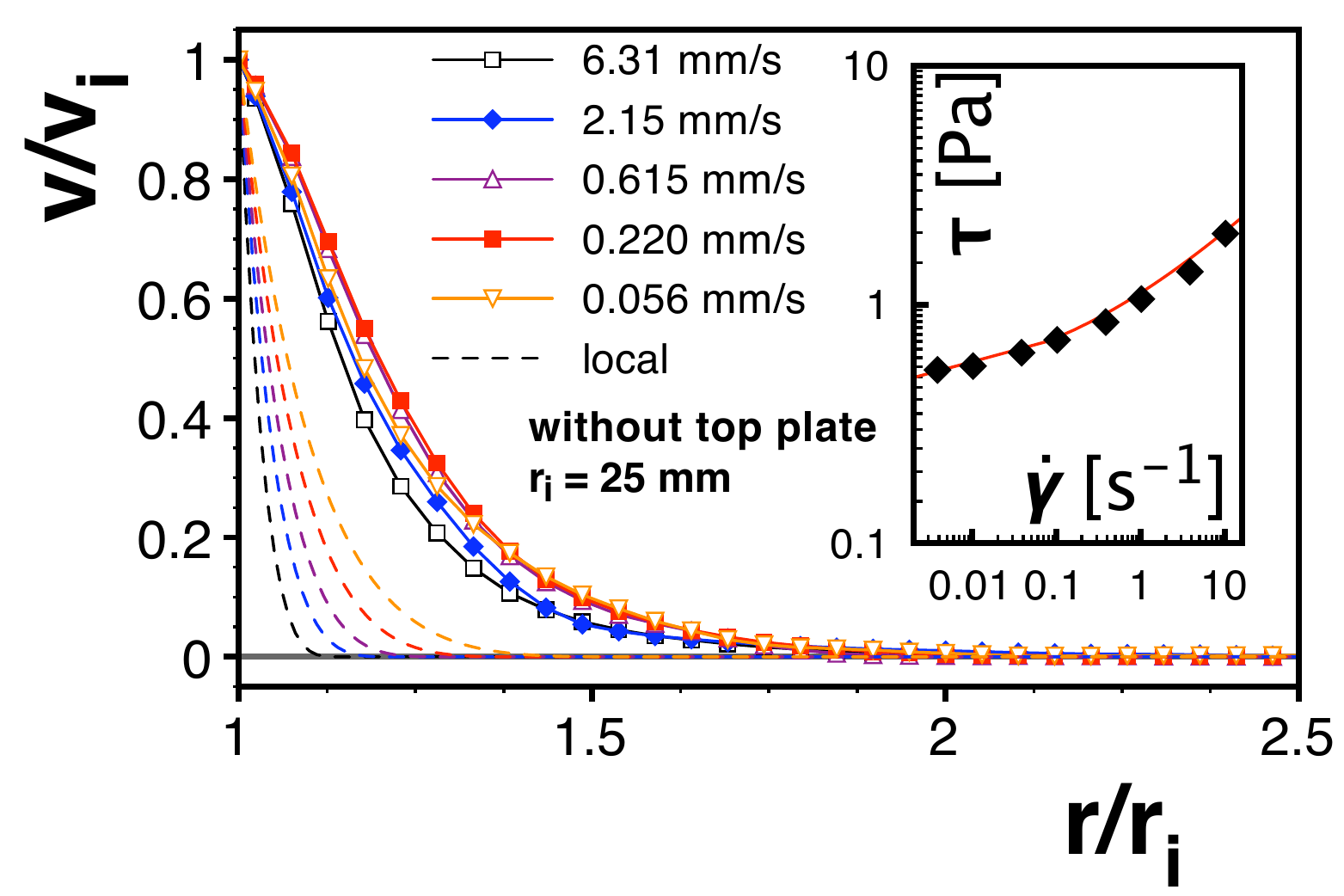}
\caption{Data from liquid-air Couette geometry with inner disk of
radius $r_i = 25$ mm driven by rheometer head. Averaged and
normalized velocity profiles for a range of driving velocities
$v_i$ at the inner disk. Dashed lines: Flow profiles for a fluid
obeying a Herschel-Bulkley constitutive relation. The HB
parameters are determined from rheometric data shown in the inset.
Black squares: Shear stress $\tau_i$ versus strain rate $\dot
\gamma$ measured at the inner disk. Red curve: Fit to a
Herschel-Bulkley constitutive relation $\tau = \tau_y + k{\dot
\gamma}^\beta$ with $\tau_y = 0.42$ Pa, $\beta = 0.36$, and $k =
0.7$ ${\rm Pa}\cdot{\rm s}^{1/\beta}$. }  \label{rheocouettetota}
\end{center}
\end{figure}

We now directly investigate the applicability of a local rheology
for the flow of a bubble raft, i.e.~in the absence of a
top plate. To do so, we have performed additional measurements by
simultaneously imaging the velocity profiles and measuring the
bulk rheometrical response of a two dimensional bubble raft in a
Couette geometry. This allows us to investigate the local rheology
of the foam in the spirit of \cite{goyon, bocquet} and connect
bulk rheometry with local measurements, as well as model
solutions.

\par
We shear the bubble raft in an Anton Paar
DSR 301 rheometer. We employ a Couette geometry, now with inner
disk and outer ring radii of $r_i = 25$~mm and $r_o = 73$~mm. We
impose five different strain rates spanning two decades and
measure the resulting averaged torque, while simultaneously
imaging the bubble motion. The measured flow profiles and rheology
are shown in Fig.~\ref{rheocouettetota}. Note that, to within
experimental scatter and similar to the Couette flows of
bubble rafts with the larger inner disk, shown in
Fig.~\ref{profiles}b, the profiles are rate independent.

An advantage of omitting the top plate is that we can determine
the stress $\tau(r)$ from the torque on the inner disk, which is
connected to the rheometer head. When there is no top
plate, the shear stress satisfies
$\frac{1}{r^2}\frac{\partial}{\partial r} (r^2 \tau) = 0$, hence
$\tau(r)=\tau_i (r_i/r)^2$. From the measured velocity profile,
the local strain rate can be obtained as $ \dot{\gamma} =
\frac{\partial v(r)}{\partial r} - \frac{v(r)}{r}$. The inset of
Fig.~\ref{rheocouettetota} plots stress versus strain rate at the
inner disk.
%This is analogous to
%the data extracted from conventional ``blind'' narrow-gap
%rheometry which, in the absence of information about the spatial
%variation of the flow within the gap, measures one shear stress
%and one strain rate.
The experimental stress-strain rate data is
consistent with a Herschel-Bulkley constitutive relation with a
yield stress of $\tau_y = 0.42$ Pa, rheological exponent of $\beta
= 0.36$, and consistency $k = 0.7$ ${\rm Pa}\cdot {\rm
s}^{1/\beta}$. This flow curve is consistent with earlier
measurements \cite{katgert, mobius}.

Equipped with a constitutive relation, we can now calculate the
expected flow profiles for given experimental parameters. We solve
for the flow profile $v(r)$ subject to the imposed driving
velocity $v(r_i) = v_i$ and no-slip boundary conditions $v(r_o)=0$
at the outer wall; results are plotted in
Fig.~\ref{rheocouettetota}. The Herschel-Bulkley flow profiles are
noticeably more shear banded than the experimental profiles and,
unlike the data, display rate dependence. Flow ceases at the point
where the stress $\tau(r)$ decays below $\tau_y$; this occurs at a
position within the gap and is clearly visible in
Fig.~\ref{rheocouettetota}. Therefore the flow profiles predicted
on the basis of the constitutive relation determined from
rheometry and imaging fail dramatically.
We now show that this departure is due to nonlocal rheology.

\section{Nonlocal effects}
\begin{figure*}[ht]
\begin{center}
\begin{minipage}[b]{0.45\linewidth}
\centering
\includegraphics[width=1\linewidth]{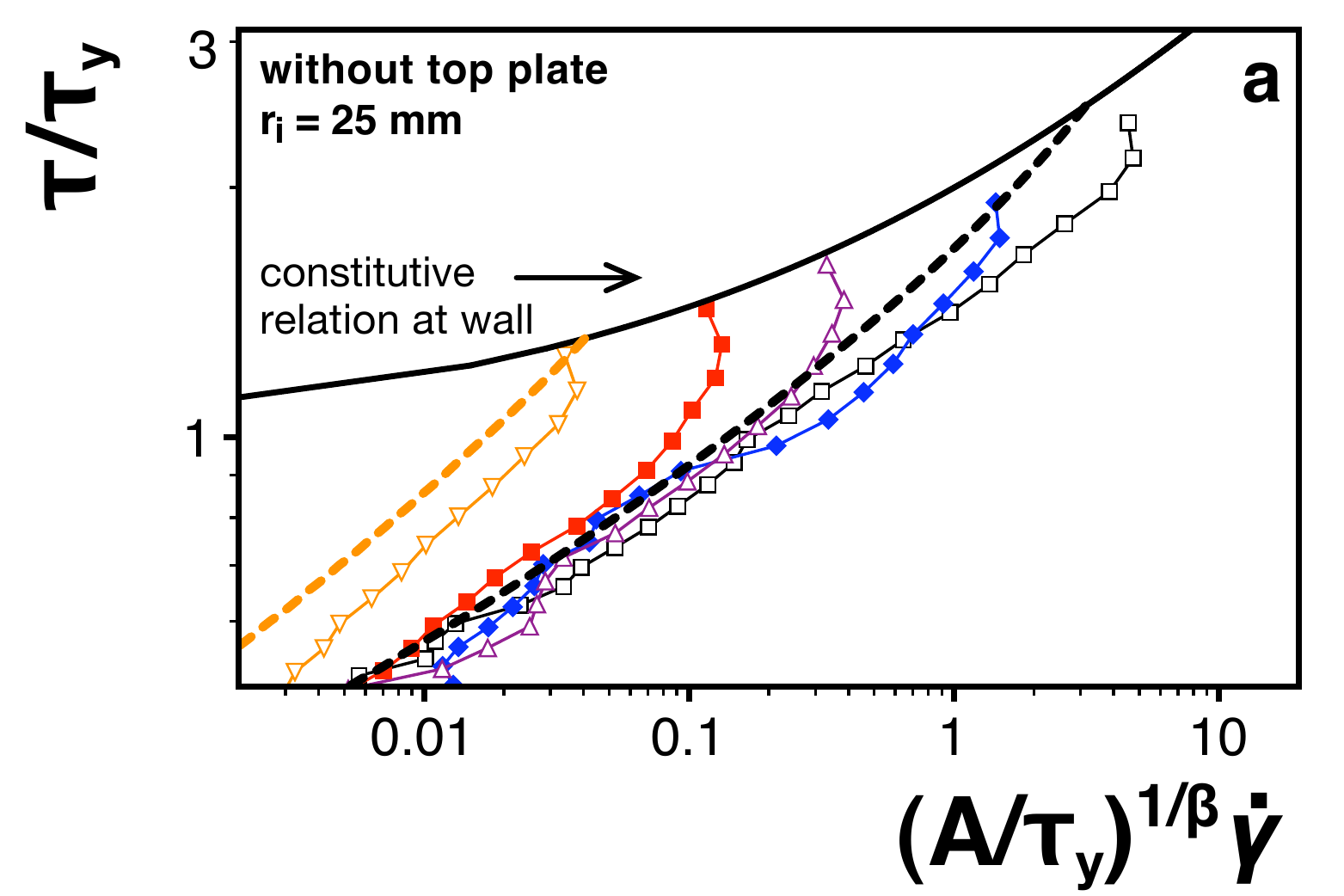}\\
\includegraphics[width=1\linewidth]{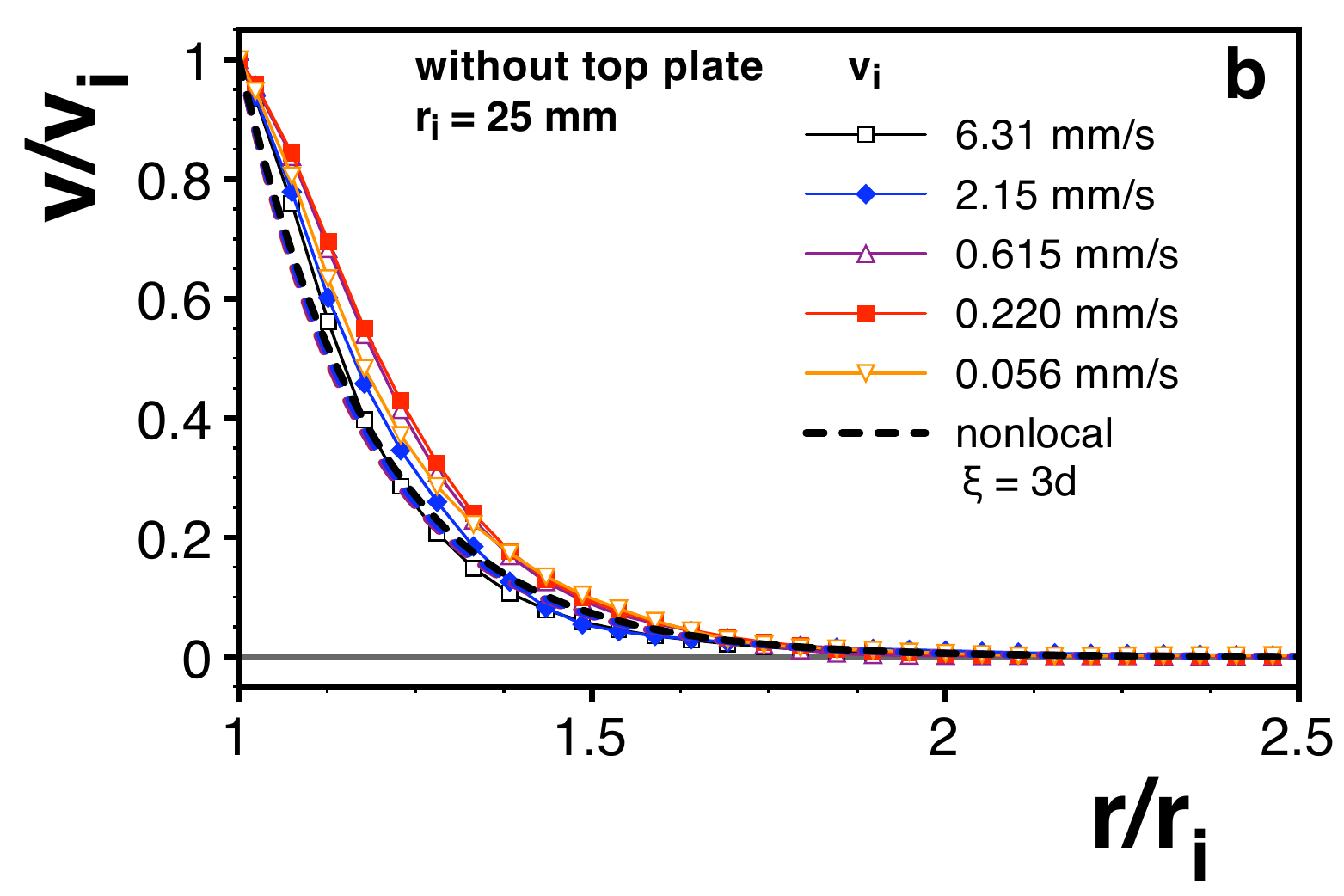}
\end{minipage}
\begin{minipage}[b]{0.45\linewidth}
\centering
\includegraphics[width=1\linewidth]{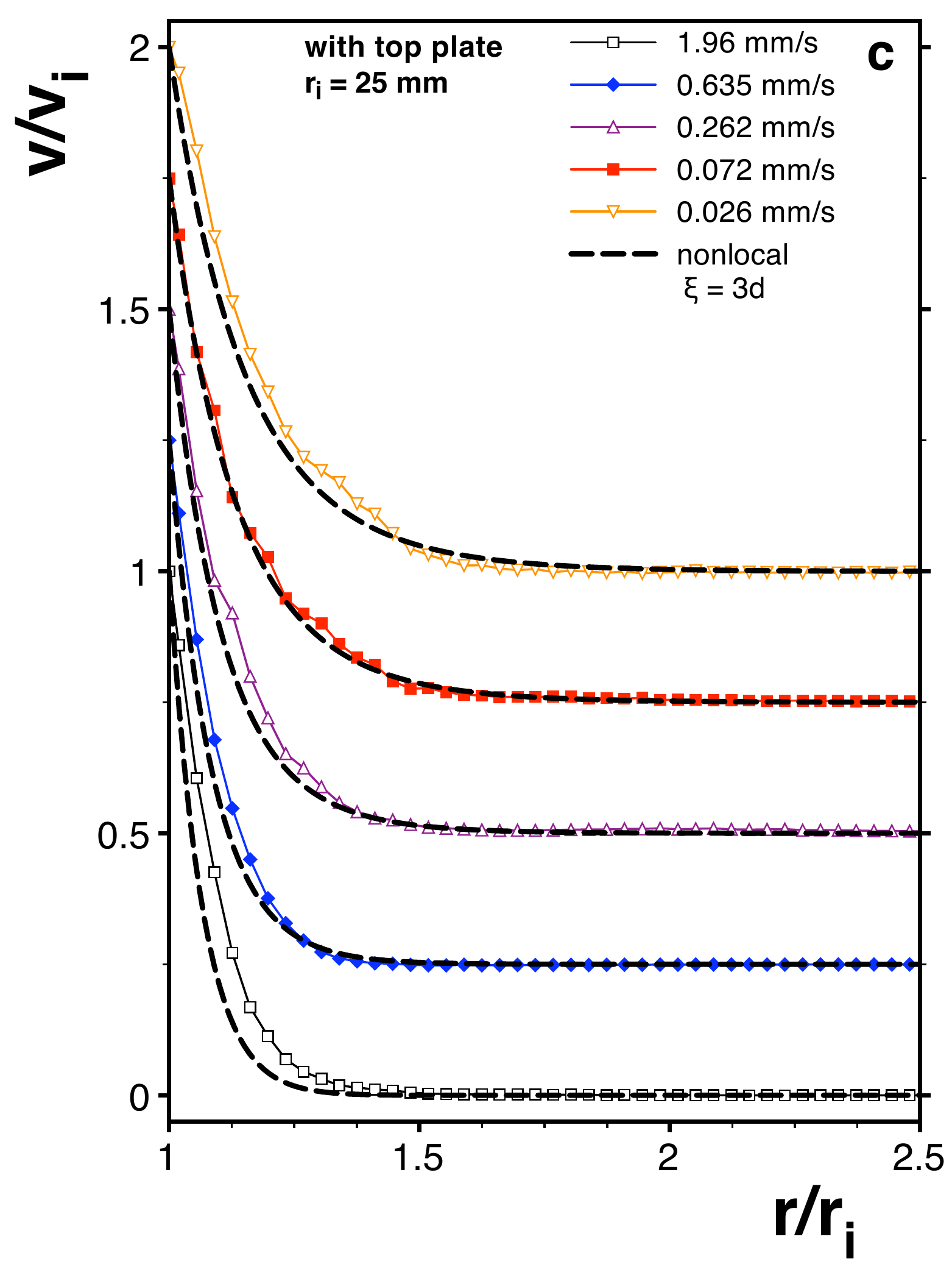}
\end{minipage}
\end{center}
\caption{Nonlocal rheology of foams. (a) Data points:
local flow curves in the Couette geometry without a top
plate. The local strain rate can be calculated from the velocity
profiles in Fig.~\ref{rheocouettetota} and the stress at the inner
disk known from the measured torque. Different colors correspond
to different driving velocities (legend as in (b)); for
a given driving velocity, different points correspond to different
positions in the gap. Solid curve: The ``wall
constitutive relation'' shown in Fig.~\ref{rheocouettetota}
(inset). Dashed curves: Predictions of the nonlocal model with
cooperativity length $\xi = 3 \langle d \rangle$. We plot only the
curves for the slowest and fastest driving rates. (b) Data points:
Identical to the data of Fig.~\ref{rheocouettetota} for flow
without a top plate. Dashed curves: Predictions of the nonlocal
model. The model yields near rate independence; the five plotted
curves lie nearly on top of one another. (c) Data points: Velocity
profiles for the system with $r_i= 25$ mm and a top plate. Dashed
curves: Predictions of the nonlocal model for the identical set of
parameters used in (b). Curves are offset for clarity. }
\label{nonlocal}
\end{figure*}

Because we access velocity profiles in addition to rheometric
data, we know the mean strain rate and mean shear stress at every
point within the gap, for the case without a top plate. It is thus
possible to make parametric plots in which the radial coordinate
is varied, as shown in Fig.~\ref{nonlocal}a.
This is equivalent to plotting a constitutive relation for each
radial coordinate within the gap. %We emphasize that all of these
%curves would collapse for a local rheology.}

%In anticipation of the model to follow,
%Fig.~\ref{nonlocal}a plots the {\em fluidity} $f(r) := {\dot
%\gamma}(r)/\tau(r)$ versus $\tau(r)$; we emphasize that this
%contains the same information as a plot of the more conventional
%pair $\tau(r)$ and ${\dot \gamma}(r)$.

Fig.~\ref{nonlocal}a demonstrates two things. First, the absence
of a collapse of the parametric plots for the five different runs
clearly shows that there is no local rheology --- for a single
given local stress, a range of local strain rates can
be obtained. If the rheology were local, all data would collapse
to a master curve, e.g.~a Herschel-Bulkley or other constitutive
relation.
%termed the bulk fluidity $f_b$, given by
%\begin{equation}
%f_b := \frac{{\dot \gamma}_{HB}}{\tau} = \frac{1}{\tau}\left(
%\frac{\tau - \tau_y}{k}\right)^\frac{1}{\beta} \Theta(\tau -
%\tau_{y}) \,.
%\end{equation}
%$\Theta(x)$ is the unit step function.
Second, we find that there
can be flow (${\dot \gamma}>0$) in the wide-gapped Couette geometry for shear
stresses {\em below} the rheometrically determined
global yield stress, $\tau(r) < \tau_{Y}$. This cannot
occur within a material that is locally described by the
Herschel-Bulkley constitutive relation.
Each parametric curve
terminates on the stress-strain rate curve of
Fig.~\ref{rheocouettetota} (inset) because the curve was measured
at the inner wheel.

We will now show that a simple nonlocal model recently developed
by Goyon {\em et al.}~can capture the flow of a bubble
raft. This nonlocal model proposes that a material's local
propensity to flow, characterized by the fluidity, can be
influenced by flow elsewhere in the material.

\section{Model}
The key ingredient of the nonlocal model is the
position-dependent inverse viscosity, or ``fluidity'', $f := {\dot \gamma}/{\tau}$,
a measure of the material's tendency to flow.
To obtain an equation for the fluidity, one can argue as follows.
A system with homogeneous stress and strain rate is 
characterized by the ``bulk fluidity''
\begin{equation}
f_b := \frac{{\dot \gamma}_{HB}}{\tau} = \frac{1}{\tau}\left(
\frac{\tau - \tilde{\tau}_y}{\tilde k}\right)^{1/\tilde\beta} \Theta(\tau -
\tilde{\tau}_{y}) \,.
\label{fbulk}
\end{equation}
$\Theta(x)$ is the unit step function. The tilde indicates that
parameters may, in principle, differ from the ``wall constitutive
relation'' of Fig.~\ref{rheocouettetota} (inset).
The key idea is that, in the presence of inhomogeneity, the system
``wants'' to achieve a fluidity $f = f_b$ everywhere, i.e.~to obey
local rheology, but is forced to pay a price for spatial
variations in $f$. Writing in one dimension for simplicity, these
ideas can be expressed very generally in integral form: $f_b(x) =
\int{\rm d}x^\prime \, K(x,x^{\,\prime}) f(x^\prime)$. The kernel
$K(x,x^\prime) = K(x-x^\prime)$ must be a symmetric function of
the distance between two points. $K$ must further satisfy
 $\int {\rm d}s\,K(s) = 1$ to recover the bulk fluidity $f_b$ in
a homogeneous system. Taylor expanding $f$ in the integrand to
second order in $s$ yields:
\begin{equation}
f_b(x) = f(x) - \xi^2 \frac{\partial^2}{\partial x^2} f(x)~.
\label{cooperativity}
\end{equation}
The parameter $\xi^2 :=  -\int {\rm d}s\, s^2 K(s) $ sets a length
scale, termed the cooperativity length, that characterizes
nonlocal effects \cite{signconvention}. Here, as in \cite{goyon},
we take it as a
fitting parameter.

Note that the diffusive term in Eq.~(\ref{cooperativity}) is in
some sense the simplest possible realization of a nonlocal model
for the fluidity. For the Couette geometry we employ cylindrical
coordinates and let $\frac{\partial^2}{\partial x^2} \rightarrow
\frac{1}{r}\frac{\partial}{\partial r}(r \frac{\partial}{\partial
r})$.

Boundary conditions on the fluidity are required: we
impose $f_i = \tau_i^{-1}[(\tau_i - \tau_y)/k]^{1/\beta}$ at the
inner wheel, with parameters from the wall constitutive relation,
and $f_o = 0$ at the outer ring. Parameters in the bulk fluidity,
Eq.~(\ref{fbulk}), must also be specified. In the experiments on
emulsions of \cite{goyon}, where it was possible to access these
parameters independently, it was found that $\tilde{\tau}_y =
\tau_y$ and $\tilde{\beta} = \beta$. For smooth walls $\tilde{k} =
k$, while for rough walls $\tilde k$ was smaller by roughly a
factor one half. For the results we present here, varying $\tilde
k$ over this range does not substantially alter the quality of
agreement between theory and experiment, both for flow curves and
velocity profiles. Hence we set $\tilde{k} = k$ for simplicity,
i.e.~we take the bulk and wall fluidities to be identical. Once
Eq.~(\ref{cooperativity}) has been solved for $f$, the velocity
profile can be integrated by assuming no slip, $v(r_o)=0$, at the
outer ring.

%Boundary conditions on the fluidity are also required: in
%Ref.~\cite{goyon} the wall fluidity is used as a fitting
%parameter. Here, for simplicity, we fix the fluidity at both
%boundaries to be the same as the bulk fluidity, $f(r_i) =
%f_b(r_i)$ and $f(r_o) = f_b(r_o)$. The latter is zero whenever
%$\tau(r_o) < \tau_y$, as is the case in all the profiles of
%Figs.~\ref{rheocouettetota}b and \ref{nonlocal}b. Finally, we
%impose no-slip boundary conditions at the outer boundary, $v(r_o)
%= 0$.

\subsection{Flows without a top plate}
For the bulk fluidity we use the Herschel-Bulkley parameters
determined by fitting to the rheometrical data in the inset of
Fig.~\ref{rheocouettetota}. We then vary $\xi$ and obtain a good
match with the data for $\xi/\langle d \rangle =3\pm0.5$. As shown
in Fig.~\ref{nonlocal}b, the flow profiles predicted by this model
are qualitatively similar to those measured, and show very little
rate dependence. In particular, the model correctly captures the
presence of flow in regions where the local stress is below the
yield criterion of the local model; this flow is induced by
cooperative effects, apparently well-captured by the nonlocal
model.

As illustrated in Fig.~\ref{nonlocal}a, this nonlocal model is in
reasonable agreement with that of the measured profiles. Both the
data and the model are roughly power law in nature,  $\tau \simeq
{\dot \gamma}^{0.2}$, although this is not exact. As a
consequence, na\"ively assuming a power-law fluid  ($\tau_y = 0$,
$\beta \approx 0.2$) captures the velocity profiles of
Fig.~\ref{rheocouettetota} reasonably well (not shown), but {\em
not} the rheometry in that figure's inset. Only the nonlocal model
resolves the apparent inconsistency between velocity profiles and
rheometry.

Note that the nonlocal model does not capture the upward bend in
the flow curves shown in Fig.~\ref{nonlocal}a, which
corresponds to regions roughly one bubble diameter from the
shearing wall where the flow gradients are small --- these are
also responsible for the ``misalignment'' of the predicted and
measured flow profiles seen in Fig.~\ref{nonlocal}b.

\subsection{Flows with a top plate}
Here we compare predictions of the nonlocal model to
flow in the presence of a top plate.
%, although the comparison is
%necessarily more qualitative. This is because rheometry cannot
%disentangle contributions to the shear stress at the inner wall
%due to bubble-bubble and bubble-wall drag, Therefore, unlike the
%case without a top plate, the appropriate Herschel-Bulkley
%relation cannot be accurately determined, though it is needed as
%input to the nonlocal model. There is additional uncertainty in
%the prefactor in $F_{bw} = c_{bw}|v|^{2/3}$, the Bretherton wall
%drag term.
%
%+ F_{bw} = 0$, where $F_{bw} \propto |v|^{2/3}$, the
%Bretherton drag force density, is present only when there is a top
%plate
We assume that the wall and bulk fluidities are unchanged from the
case without a top plate and take $\xi/ \langle d \rangle = 3$,
the same parameters used in Figs.~\ref{nonlocal}a and b. We
include a Bretherton wall drag $F_{bw} = c_{bw}|v|^{2/3}$, where
$c_{bw} = 2.7\times10^5$ ${\rm Pa}\cdot{\rm
m}^{-\frac{5}{3}}\cdot{\rm s}^\frac{2}{3}$ has been determined
independently \cite{katgert}. In Fig.~\ref{nonlocal}c we compare
flow profiles from the experiment and model.  The model
predictions are in surprisingly good agreement with experimental
results. Rate dependence emerges, and the flow profiles are more
localized than in the case absent a top plate.

\section{Outlook}

We probed the flow profiles and rheology of two-dimensional foams
in Couette geometries, both for foams squeezed below a top plate
and for 
bubble rafts. Consistent with earlier
experimental results in a linear geometry \cite{katgert}, flows
below a top plate are strongly rate dependent. While for
bubble rafts in a linear geometry one expects and finds
rate independence \cite{denninpre73,janiaud}, this is not to be
expected in the Couette geometry, since a combination of a finite
yield stress and a radial decay of the shear stress suggests a
rate dependent location in the cell where the flow ceases
\cite{denninpre74,denninreview}. Here we do not observe this
effect: the velocity profiles for a wide range of strain rates
collapse.

For bubble rafts, the local and global rheology do not
match, and in particular the foam flows in regions where the
stress is below the global yield stress.  We can fit the flow
profiles of both the bubble rafts and the confined
foams by a nonlocal model that extends the measured
Herschel-Bulkley rheology with an empirically determined length
scale that captures the nonlocality.

We note that the effect of nonlocality for confined foams is
not as strong as it is for the freely floating foams. Moreover, we
suggest that the nonlocal effect would be even less important in
the linear geometry, where a local model was found to capture the
flow profiles \cite{katgert}. Underlying this is that even the
local model predicts no abrupt cessation of flow in linear
geometry  --- and it is near such regions that the flow  profiles
are most sensitive to nonlocal effects.

One open question remains. If we assume the Herschel-Bulkley
relation determined in the system with inner radius $r_i=25$ mm
also describes bubble raft in the system with $r_i =
105$ mm, we can attempt to describe the velocity profiles in the
large system using the nonlocal model. We obtain good agreement
for a choice of cooperativity length $\xi / \langle d \rangle
\approx 10$, significantly larger than that in the small system.
We suggest that the radius of curvature may influence the
cooperativity length in a way not captured in the model as
presented here.

\acknowledgments The authors wish to thank Anna Baas for
experimental assistance and Jeroen Mesman for technical
brilliance. It is a pleasure to thank Lyderic Bocquet and Wim van
Saarloos for enlightening discussions. GK, BPT and MEM acknowledge
funding by Stichting FOM.


\begin{thebibliography}{0}
\bibitem{signconvention}
The integral $\int {\rm d}s\, s^2 K(s)$ can be positive or
negative. Real values of $\xi$ correspond to the negative case.
Imaginary $\xi$ tends to produce non-monotonic velocity profiles.
\end{thebibliography}
\end{document}